\pdfoutput=1
\documentclass{article}
\usepackage{graphicx}
\begin{document}
\title{\textbf{The PADME experiment at Laboratori Nazionali di Frascati}}
\author{
Gabriele Piperno for the PADME collaboration      \\
\footnotesize{\em INFN Laboratori Nazionali di Frascati, Via E. Fermi, 40 -- I-00044 Frascati (Rome), Italy}
\\
}
\maketitle
\baselineskip=11.6pt

\begin{abstract}

The PADME experiment will search for the invisible decay of Dark Photons
produced in interactions of positron from the DA$\Phi$NE Linac on
a target. The collaboration aims at reaching a sensitivity of $\sim10^{-3}$
on the coupling constant for values of Dark Photon masses up to $23.7\,\mbox{MeV}$.

\end{abstract}
\baselineskip=14pt
%

\section{Introduction}

The problem of the elusiveness of the Dark Matter (DM) can be solved
speculating that it interacts with particles and gauge fields of the
Standard Model (SM) only by means of portals that connect our world
to the dark sector. The simplest model adds a U(1) symmetry and
its vector boson $A'$: SM particles are neutral under this symmetry,
while the new boson couples to the SM with an effective charge $\varepsilon e$
and for this reason it is called Dark Photon (DP) \cite{DP U(1),U(1)_2}.

In addition, it has been pointed that the existence of an $A'$ with
a mass $m_{A'}$ in the range $[1\,\mbox{MeV},1\,\mbox{GeV}]$ and
a coupling constant $\varepsilon\sim10^{-3}$, might be responsible
for the discrepancy currently observed between the theoretical expectation,
based on the SM, and the measurement of the muon anomalous magnetic
moment $(g-2)_{\mu}$ \cite{muon g-2}.

\begin{figure}
\begin{centering}
\includegraphics[height=5cm]{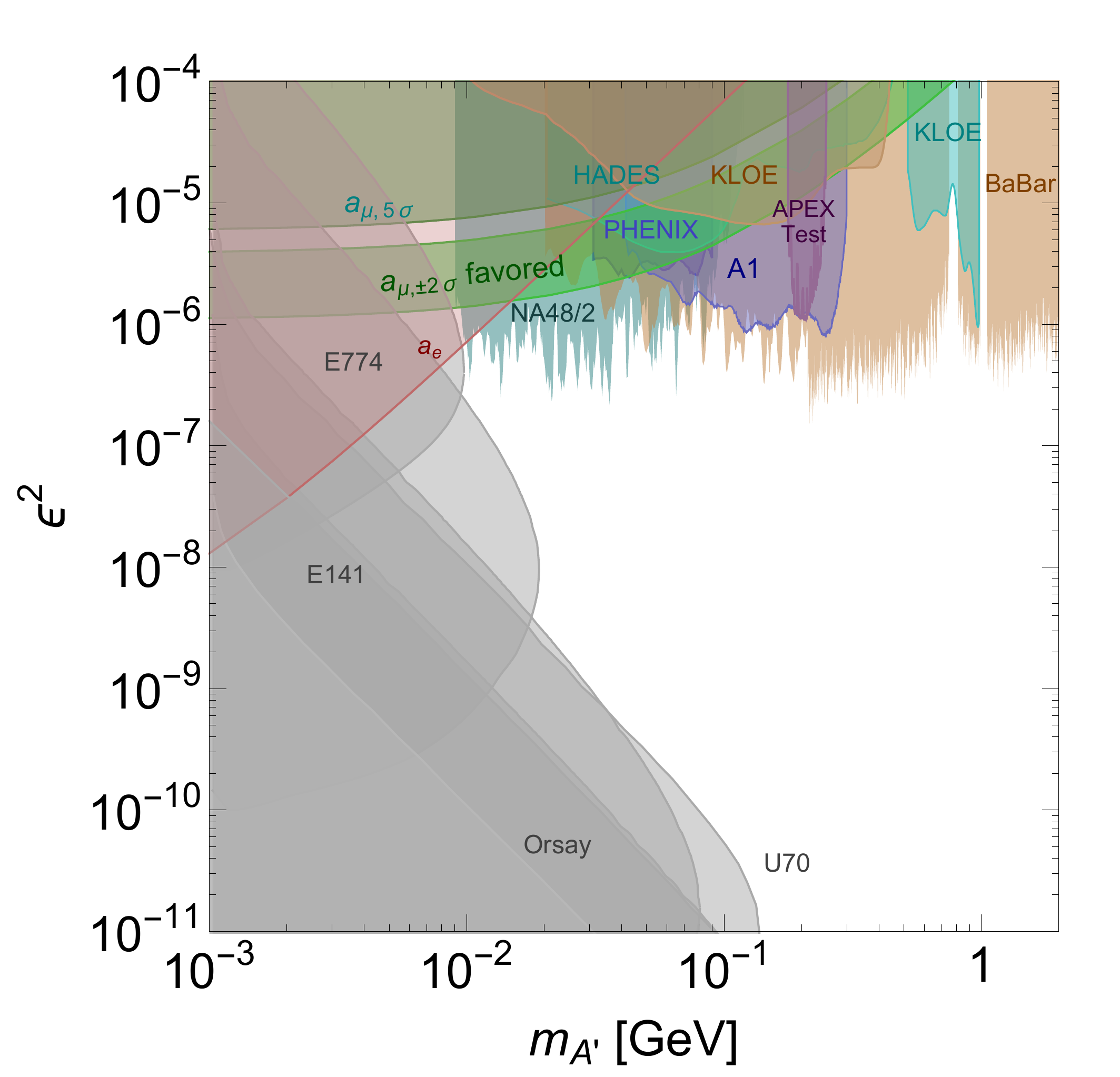}\includegraphics[height=5cm]{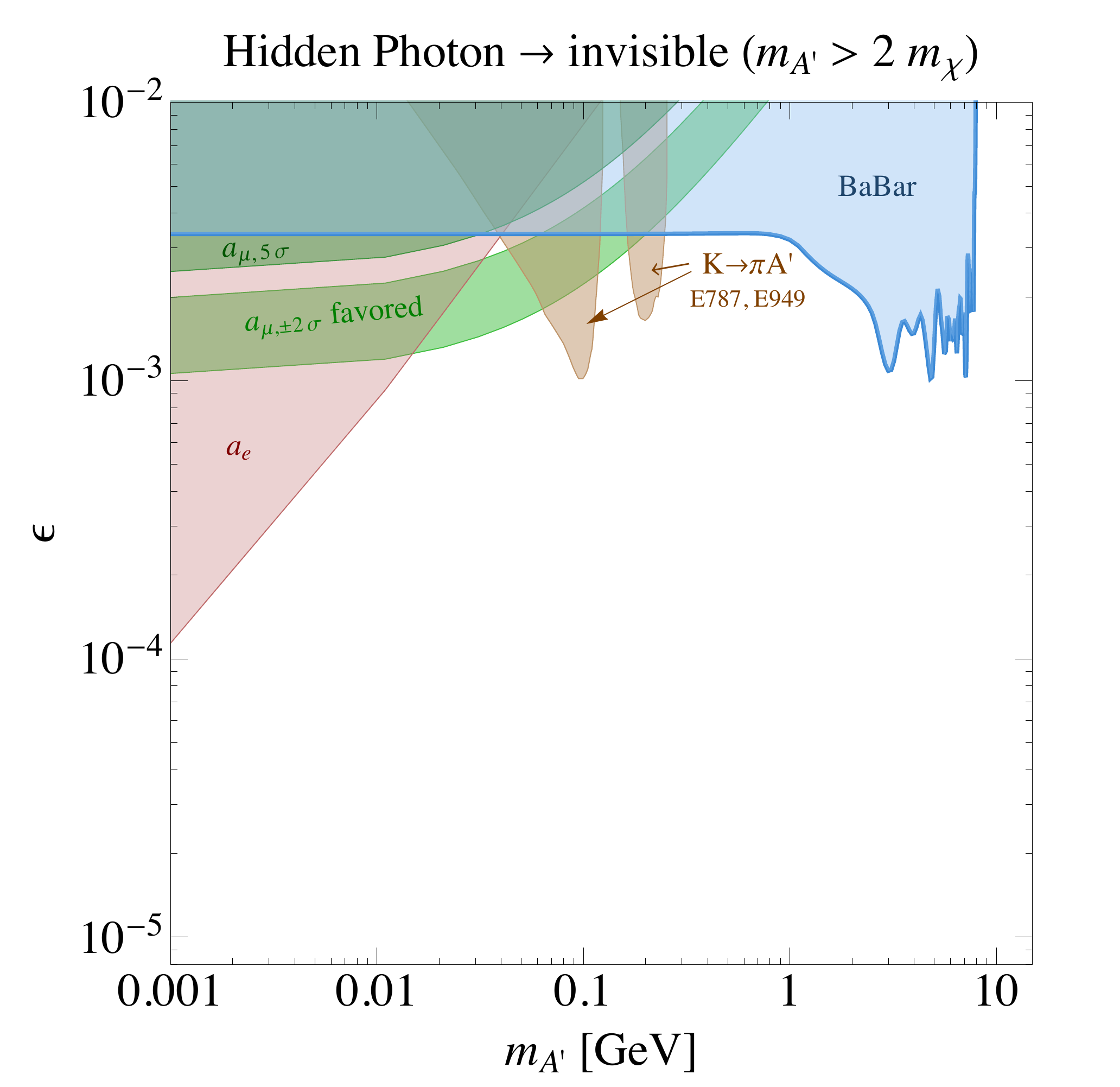}
\par\end{centering}

\caption{\label{fig:Search Status}Current DP search status: on the left for
visible decays (adapted from \cite{visible status}), on the right
for the invisible ones (adapted from \cite{invisible status}). Typical
DP exclusion plot have the $A'$ mass on the x-axis and the coupling
constant (squared) on the y-axis. In both cases the $2\sigma$ anomalous
muon magnetic moment favored band is indicated.}
\end{figure}

If there are no particles in the hidden sector with mass smaller than
one half of $A'$, it can only have SM decays (visible decays). Currently,
the region of the $\varepsilon$, $m_{A'}$ plane favored by $(g-2)_{\mu}$
discrepancy, is excluded for an $A'$ decaying into SM (see Fig.\ref{fig:Search Status}
left).

In the most general case the $A'$ can decay into DM (invisible decays).
In this scenario, there are still unexplored regions in the $(g-2)_{\mu}$
favored band, as shown in Fig.\ref{fig:Search Status} right.

A comprehensive overview of the experimental programs of this field
is presented in \cite{DP search}.

\section{The PADME experiment}

The PADME (Positron Annihilation into Dark Mediator Experiment) experiment
is designed to detect invisible decaying DPs that are produced in
the reaction $e^{+}\, e^{-}\rightarrow A'\,\gamma$, where the $e^{+}$
are accelerated from the DA$\Phi$NE to $550\,\mbox{MeV}$ and the
$e^{-}$ belongs to a fixed target \cite{PADME,PADME 2}.

\subsection{The Frascati Beam Test Facility}

PADME will be hosted in the newly redesigned hall of the Beam Test
Facility (BTF), a transfer-line from the DA$\Phi$NE linac of the
Laboratori Nazionali di Frascati (LNF) \cite{BTF}. BTF is able to
provide up to $50\,\mbox{bunches/s}$ with a maximum energy of $550$
and $800\,\mbox{MeV}$, for positrons and electrons respectively,
and with duration (at constant intensity) from $1.5$ to $40\,\mbox{ns}$.
The energy spread is $0.5\%$, while the beam spot size can vary by
orders of magnitude: {[}$0.5,25]\,\mbox{mm}\,(\mbox{vertical})\,\times\,[0.6,55]\,\mbox{mm}\,(\mbox{horizontal})$.
The number of particles that can be provided per bunch goes from $1$
to $10^{10}$.

\subsection{The detector}

The detector is designed to identify events with a single photon emerging
from the $e^{+}/e^{-}$ annhilation and to measure the missing squared
invariant mass of the final state, by exploiting energy-momentum conservation
and the fully constrained initial state: $e^{+}$ beam (known momentum
and position) on an active fixed target. The $A'$ squared invariant
mass $M_{miss}^{2}$ can be estimated as:

\[
M_{miss}^{2}=\left(\vec{P}_{e^{-}}+\vec{P}_{beam}-\vec{P}_{\gamma}\right)^{2},
\]

where $\vec{P}_{e^{-}}=\vec{0}$ and $P_{beam}=550\,\mbox{MeV}$ along
the initial beam direction, are the $e^{-}$ and the $e^{+}$ momentum
respectively and $\vec{P}_{\gamma}$ is the photon final state.

\medskip{}

\begin{figure}
\begin{centering}
\includegraphics[height=4.8cm]{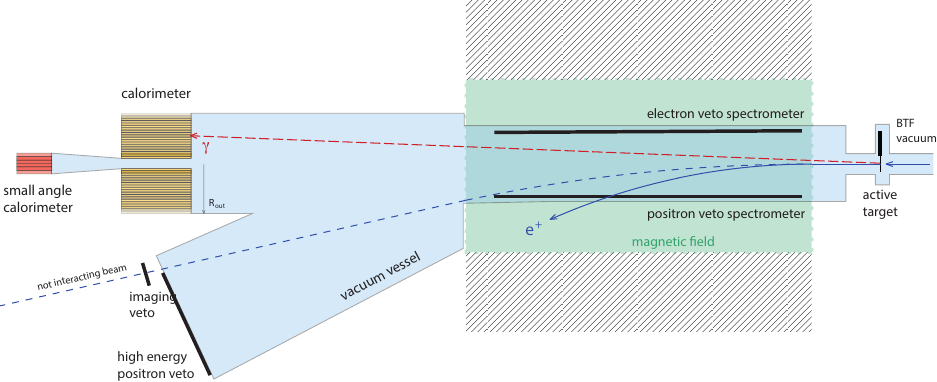}
\par\end{centering}

\caption{\label{fig:Detector}PADME detector layout. From right to left: the
active target, the $e^{+}e^{-}$ vetoes inside the magnetic dipole,
the high energy $e^{+}$ veto near the exhaust beam exit, the ECAL
and the SAC. The distance between the ECAL and the target is $3\,\mbox{m}$.}
\end{figure}

The detector, shown Fig.\ref{fig:Detector}, consists of different
components \cite{PADME}:
\begin{itemize}
\item Diamond active target. It allows to measure the beam intensity and
position (precision of $\approx5\,\mbox{mm}$) by means of graphite
perpendicular strips. The low Z of diamond is needed to reduce the
bremsstrahlung process. The area is $2\times2\,\mbox{cm}^{2}$ and
the small thickness ($50\,\mu\mbox{m}$ or $100\,\mu\mbox{m}$) is
to reduce the probability of $e^{+}$ multiple interactions.
\item Dipole magnet. Located $20\,\mbox{cm}$ after the target, it is designed
to deflect exhaust beam out of the detector and send the positrons
that lost part of their energy (mainly through bremsstrahlung) towards
the vetoes. The field is $0.5\,\mbox{T}$ over a gap of $23\,\mbox{cm}$
for $1\,\mbox{m}$ of length.
\item Positrons/electrons veto. It is divided into two parts: one inside
the dipole for positrons and electrons and one, near the beam exit,
for high energy positrons that lost only a small part of their energy,
typically for bremsstrahlung. It is composed of $1\times1\times16\,\mbox{cm}^{3}$
bars of plastic scintillators. The arrays inside the magnet are $\approx1\,\mbox{m}$
long, while the high energy positron one is $\approx0.5\,\mbox{m}$
long.
\item Electromagnetic calorimeter (ECAL). Made of $616$ $2\times2\times22\,\mbox{cm}^{3}$
BGO crystals and placed at $3\,\mbox{m}$ from the target. Energy
resolution is foreseen to be $\sim\frac{(1-2)\%}{\sqrt{E}}$. The
shape is cylindrical ($30\,\mbox{cm}$ radius) with a central hole
(a square of $10\,\mbox{cm}$ side) to allow the bremsstrahlung radiation
to pass and impinge on the Small Angle Calorimeter. This is necessary
because of the BGO decay time of $300\,\mbox{ns}$: the ECAL would
be continuously ``blinded'' by the bremsstrahlung rate. The angular
coverage is $(20,93)\,\mbox{mrad}$.
\item Small Angle Calorimeter (SAC). It consists of $49$ $2\times2\times20\,\mbox{cm}^{3}$
lead glass SF57 and its goal is to veto events with a bremsstrahlung
photon. The lead glass decay time of $4\,\mbox{ns}$ makes it a good
candidate for this task, being fast enough for the expected rate.
The angular coverage is $(0,20)\,\mbox{mrad}$.
\end{itemize}

Hence the DP signature is a single $\gamma$ in the ECAL and no particles
in the vetoes. Being $E_{beam}=550\,\mbox{MeV}$, the largest $A'$
reachable mass is $23.7\,\mbox{MeV}$.

\subsection{Backgrounds and sensitivity}

\begin{figure}
\begin{centering}
\includegraphics[height=4.5cm]{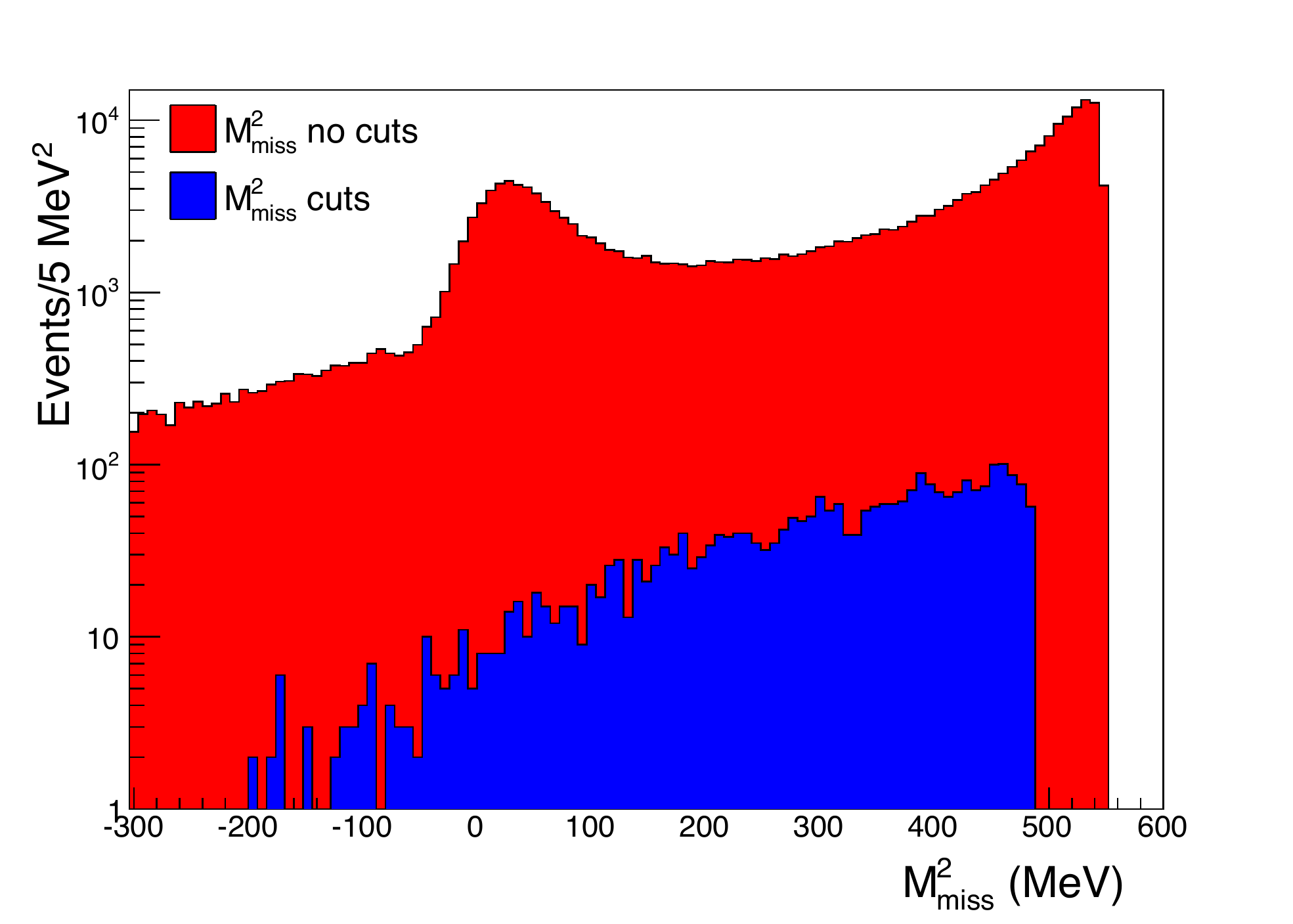}
\par\end{centering}

\caption{\label{fig:Backgrounds}Background before (red) and after (blue) good
events selection.}
\end{figure}

The SM physical processes that take place when the $e^{+}$ beam hits
the target are: bremsstrahlung and $e^{+}/e^{-}$ annihilation in
2 or 3 $\gamma$s.\cite{PADME} The probability that they mimic a DP production
event can be reduced through an optimization of the ECAL geometry
and granularity and of the system of vetoes. The beam intensity plays
an important role through the pileup: clusters cannot be resolved
in time by the calorimeter if they are temporally too close each other.\cite{PADME}
Fig.\ref{fig:Backgrounds} shows the background reduction obtained
requiring only one cluster in the ECAL, no hits in vetoes, no $\gamma$s
in the SAC with energy $>50\,\mbox{MeV}$ and an energy of the cluster
in a range optimized depending on $m_{A'}$.

\begin{figure}
\begin{centering}
\includegraphics[height=5cm]{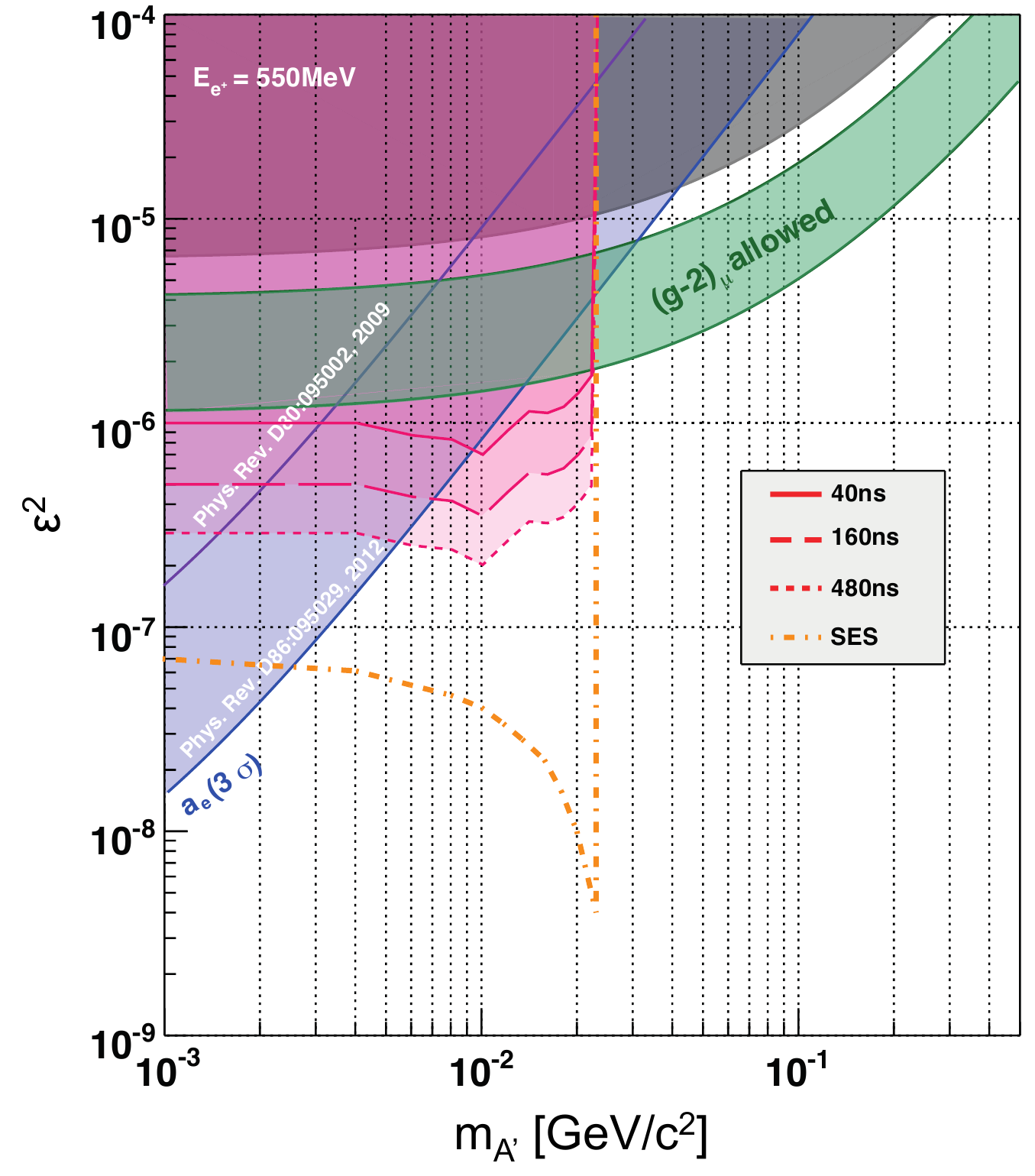}
\par\end{centering}

\caption{\label{fig:Sensitivity}PADME sensitivity to $A'\rightarrow invisible$.
Increasing bunch length it is possible to explore smaller $\varepsilon$.
SES refers to sensitivity in absence of background.}
\end{figure}

The DP sensitivity calculation is based on $2.5\cdot10^{10}$ GEANT4
simulated $550\,\mbox{MeV}$ positrons on target extrapolated to $10^{13}$
$e^{+}$. This number of particles can be obtained running PADME for
$2\,\mbox{y}$ at $60\%$ efficiency with $5000$ $e^{+}$ per bunch
($40\,\mbox{ns}$) at a repetition rate of $50\,\mbox{Hz}$. The obtained
result for a DP decaying to invisible particles is shown in Fig.\ref{fig:Sensitivity}
for different bunch durations: favored $(g-2)_{\mu}$ region can be
explored in a model independent way (the only hypothesis on the DP
is the coupling to leptons) up to masses of $23.7\,\mbox{MeV}$.\cite{PADME} Single
Event Sensitivity (SES) refers to the sensitivity in absence of background.

\section{Conclusions}

Theoretical models with a DP provide a solution to the DM puzzle.
Additionally a DP with mass in the $[1\,\mbox{MeV},1\,\mbox{GeV}]$
interval and coupling constant $\varepsilon\sim10^{-3}$, can justify
the muon anomalous magnetic moment discrepancy.

PADME will perform a model independent search for an invisible decaying
DP, using the accelerator complex present at the LNF. The collaboration
aims at reaching a sensitivity on $\varepsilon$ of $\sim10^{-3}$
for DP with masses up to $23.7\,\mbox{MeV}$.



%
\end{document}